# The Dynamical Mechanism of the Aharonov-Bohm Effect

R. F. Wang

*Department of Physics, Beijing Jiaotong University,*

*Beijing, 100044,    People's Republic of China*

*E-mail: aboud56789@163.com*

**abstract**

In this paper, it is emphasized that the dynamical cause for the A-B effect is the superimposed energy between the magnetic field produced by the moving charges and that in the solenoid, instead of the existence of the vector potential $\mathbf{A}$. If such a superposition between the magnetic fields can be eliminated, the A-B effect should not be observed any more. To verify this viewpoint, a new experimental method using a SQUID is suggested in this paper.

## 1    Introduction

In 1959, Y. Aharonov and D. Bohm[1] predicted that a phase difference between two electron beams is produced proportion to the magnetic flux enclosed by their paths, even though they always move in the region where the magnetic field is zero. This effect is difficult to be understood in classical physics and has led to extensive debate. Some authors asserted that the A-B effect doesn't exist[2]; Some authors[3] explained this effect from the standpoint that electrons were affected by inevitable leakage magnetic fields from the finite solenoid in the experiments[4]. Both views were removed by the toroidal experiments of Tonomura[5] which allowed very little stray magnetic flux and indicated that the A-B effect does exist.

Other authors suggested that the Aharonov-Bohm shift might be based upon a classical lag effect[6][7]. In their view, the passing charge experiences a net force due to interaction with the solenoid. But this view still needs to be verified by experiments[8], for whether the force acting on the passing charge exists is an open question[9][10].

Since the experiments by Tonomura[5], no one questions the existence of the Aharonov-Bohm effect, and this effect has become a standard part of the quantum mechanics textbook. The current view is: the Aharonov-Bohm effect represents a new quantum topological effect, and the vector potential **A** can result in some observable phenomena in quantum mechanics, though it is only a mathematical field in classical physics[1]. But in our view, this view doesn't reveal the physical nature of the Aharonov-Bohm phase shift.

In our view, the concept of 'force' plays a key role in classical mechanics. All the phenomena can be explained by analyzing the forces acting on the objects. But in quantum mechanics, the concept of 'energy' becomes the most important, in contrast, the concept of 'force' disappears. All the phenomena in quantum mechanics are determined by the Hamiltonian, which represents the energy of the system. As the A-B effect being a typical quantum phenomenon, we should explain this effect with the standpoint of energy, instead of the standpoint of topology or classical electromagnetism.

In this paper, We will point out that the A-B effect is due to the superposition between the magnetic field produced by the moving charges and that in the solenoid, but not the vector potential **A**. If such a superposition can be eliminated, the A-B effect will disappear, even though the vector potential still exists in the space.

A similar proposal was ever offered by Erlichson[11]. To verify this proposal, an experimental method similar to that used by Tonomura was suggested too. But as showed in this paper, such an experimental method cannot judge whether the A-B effect results from vector potential **A** or from the superimposed energy between the magnetic fields. To verify that the A-B effect is due to the superimposed energy but not the vector potential **A**, we will first provide a theory proof for this conclusion and then provide a feasible experimental method to test it.

## 2  The dynmatical analysis about the A-B effect

We suppose that the static magnetic field created by the solenoid is denoted by $\mathbf{B}_0$, which only exists in the region $\Omega$ enclosed by the solenoid; and the magnetic field produced by a moving charge $q$ is denoted by $\mathbf{B}_1$, which is determined by the following formula[12]:

$$\mathbf{B}_1(\mathbf{r}) = \frac{\mu_0}{4\pi} \frac{q\mathbf{v} \times (\mathbf{r} - \mathbf{x})}{|\mathbf{r} - \mathbf{x}|^3} \tag{1}$$

where, $\mathbf{B}_1(\mathbf{r})$ denotes the magnetic induction at the point $\mathbf{r}$, $\mathbf{v}$ and $\mathbf{x}$ are the velocity and position of the charge $q$.

The magnetic field $\mathbf{B}_1(\mathbf{r})$ can superposition with the field $\mathbf{B}_0$. The mutually superimposed energy $W'$ between these two magnetic fields is:

$$W' = \int \frac{1}{\mu_0} \mathbf{B}_0 \cdot \mathbf{B}_1 dr^3 = \frac{1}{4\pi} \int_\Omega \mathbf{B}_0(\mathbf{r}) \cdot \frac{q\mathbf{v} \times (\mathbf{r} - \mathbf{x})}{|\mathbf{r} - \mathbf{x}|^3} dr^3$$

$$= \frac{1}{4\pi} \int_\Omega \frac{[\mathbf{B}_0(\mathbf{r}) \times (\mathbf{x} - \mathbf{r})] \cdot q\mathbf{v}}{|\mathbf{x} - \mathbf{r}|^3} dr^3 = \mathbf{A}(\mathbf{x}) \cdot q\mathbf{v} \tag{2}$$

where, $\mathbf{x}$ still denotes the position of the moving charge, and

$$\mathbf{A}(\mathbf{x}) = \frac{1}{4\pi} \int_\Omega \frac{[\mathbf{B}_0(\mathbf{r}) \times (\mathbf{x} - \mathbf{r})]}{|\mathbf{x} - \mathbf{r}|^3} dr^3 \tag{3}$$

Obviously, $\nabla \times \mathbf{A} = \mathbf{B}_0$ and $\nabla \cdot \mathbf{A} = 0$, *i.e.* $\mathbf{A}(\mathbf{x})$ is a vector potential field describing the magnetic field $\mathbf{B}_0(\mathbf{r})$ and satisfies the Coulomb gauge. Here, the vector potential $\mathbf{A}(\mathbf{x})$ is determined uniquely and any gauge transformation is forbidden, for if we add to $\mathbf{A}(\mathbf{x})$ the gradient of a scalar function $\chi(\mathbf{x})$ and form $\mathbf{A}'(\mathbf{x}) = \mathbf{A}(\mathbf{x}) + \nabla\chi(\mathbf{x})$, then $\mathbf{A}'(\mathbf{x}) \cdot q\mathbf{v}$ doesn't equal the mutually superimposed energy $W'$. For this reason, all the vector potentials in this paper are determined by the formula (3).

As showed in Fig.1, a coherent electron beam splits into two beams at F. One beam bypasses the solenoid through the side C, and the other through the side D. Then, they combine to interfere at E. The direction of the magnetic field in the solenoid is perpendicular to the

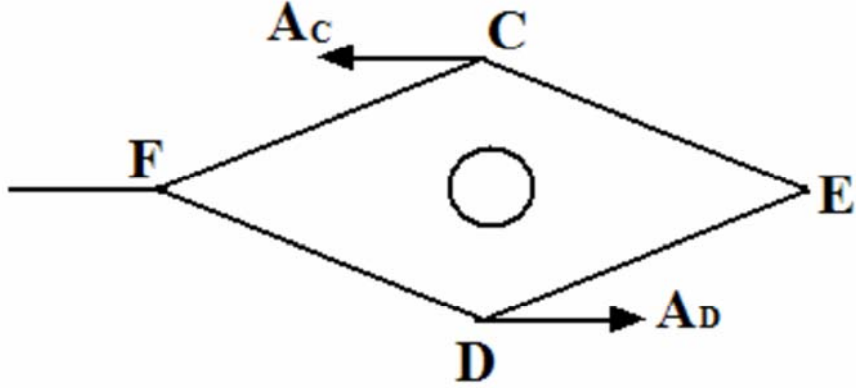

Fig1, The dynmatical analysis about the A-B effect.

paper and points outside. $\mathbf{A}_C$ and $\mathbf{A}_D$ are the vector potentials at the points C and D. According to the exp (3), $\mathbf{A}_C = -\mathbf{A}_D$. Suppose $\mathbf{v}_C$ and $\mathbf{v}_D$ are the velocities of the electrons at the points C and D, obviously, $\mathbf{v}_C = \mathbf{v}_D$. So, when an electron passes by the point C, the superimposed energy $W'_C$ between the magnetic field produced by this electron and the static magnetic field in the solenoid is equal to $\mathbf{A}_C \cdot (-e)\mathbf{v}_C$, where $-e$ is the charge of an electron. Similarly, when an electron passes by the point D, the superimposed energy $W'_D$ equals to $\mathbf{A}_D \cdot (-e)\mathbf{v}_D$, obviously, $W'_C = -W'_D$. This difference in superimposed energy inevitably results in the difference of the wave functions at the points C and D. Therefore, the patter of interference at E should change with the magnetic flux in the solenoid[13], *i.e.* the A-B effect. So, the superimposed energy can be considered as the dynamical cause for the A-B effect. If we have a way to eliminate the superimposed energy, the A-B effect should disappear.

To verify this deduction, a simple method is to coat a superconductor cylinder outside the solenoid. When the superconductor cylinder comes into the superconducting state, due to the Messiner effect, it can completely shield the magnetic field due to the moving electrons out of the cylinder. Consequently, the magnetic field produced by the moving electrons cannot superimpose with the static magnetic field in the solenoid, and the superimposed energy between them becomes zero. According to the analysis above, the A-B effect should not be observed any more. But this conclusion is contradictory to the experiments by A. Tonomura *et al* [5]. In their experiments, a tiny toroidal magnet covered entirely with a superconductor layer and further with a copper layer is fabricated. The thickness of the superconductor layer (about $250nm$) is much lager than its magnetic penetration depth. When the temperature is below the critical temperature of the superconductor layer, the magnetic flux $\phi_S$ enclosed within the superconducting layer is quantized, *i.e.* $\phi_S = nh/2e$ [14][15].

The A-B effect asserts that the relative phase shift $\Delta\varphi$ between two electron beams is proportion to the magnetic flux $\phi$ enclosed by their paths:

$$\Delta\varphi = 2\pi \frac{\phi}{h/e} \quad (4)$$

If $\phi_S = (2n+1)\left(\frac{h}{2e}\right)$, *i.e.* the phase shift $\Delta\varphi$ between two electron beams equals to $(2n+1)\pi$, each interference fringe inside the ring should lie just in the middle of two fringes outside the ring; If $\phi_S = 2n\left(\frac{h}{2e}\right)$, *i.e.* the phase shift $\Delta\varphi$ equals to $2n\pi$, the interference fringes inside the ring should align with the fringes outside the ring. Both cases have been observed in their experiments. That is to say, though the solenoid is coated by a superconducting cylinder, the magnetic flux enclosed by the solenoid can still affect the quantum phase of the electrons passing outside the cylinder. Obviously, these experimental results are contradictory to our analysis about the A-B effect. But the further analysis can show that our viewpoint is not wrong. The cause for this contradiction is that the superconducting layer in their experiments can only confine the magnetic flux within it, but cannot shield the magnetic field produced by the electron beams.

In their experiments, an electron beam can be regarded as a system of successive wave packets, and the coherent length of each wave packet is about $3 \sim 5\mu m$ [16]. The wave packets are separated with each other and the quantum interference that occurs in forming an electron hologram involves one electron at a time. Each wave packet produce a magnetic field at the tiny toroidal magnet. According to the exp (1), the magnitude of this magnetic field is inversely proportion to the square of the distance. So, the magnetic field at the tiny magnet forms a pulse, and its maximum is produced when the wave packet passes through the plane where the tiny magnet lies. The time width of this pulse equals to the length of the wave packet $\Delta l$ divided by its velocity $v$, i.e. $\Delta t \approx \Delta l / v$ In their experiments, the velocity of the electron wave packets is about $2 \times 10^8 m/s$ so, $\Delta t \approx 2 \times 10^{-14} s$. If the Fourier transform is introduced, the main frequency of this pulse is about $\nu \approx 5 \times 10^{13} Hz$, i.e. $h\nu \approx 2 \times 10^{-2} eV$ which is much larger than the energy gap in the Nb film (about $3 \times 10^{-3} eV$). Obviously, the Nb film cannot shield the magnetic field variation with so high frequency. Therefore, the magnetic field produced by the moving electron can still penetrate the superconducting film and superposition with the static magnetic field in the tiny magnet. Just for this reason, the A-B effect was observed in their experiments. If one wants to verify what is the dynamical cause for the A-B effect, the vector potential **A** or the superimposed energy between the magnetic fields? a new experiment have to be designed, in which the superconducting film must be able to completely shield the magnetic field produced by the moving charges. For this purpose, a new theorem needs showing at first.

## 3  A new theorem

Suppose there is a straight and infinitely long solenoid, outside which a infinitely long superconductor cylinder coats. The depth of this cylinder is much larger than its magnetic penetration depth. A charge system *G* lies outside the cylinder, and its frequency $\nu$ is far smaller than $\Delta / h$ ( where $\Delta$ is the energy gap of the superconductor). Then, when the cylinder comes into the superconducting state, the magnetic field enclosed by the superconducting cylinder cannot have any effect on the charge system *G*.

For simplicity, we assume that the moving charges in the system *G* are only electrons. Obviously, the current distribution in the system *G* will produce a magnetic field around itself, and induce a shielding current distribution on the outer surface of the superconductor cylinder ensuring that the magnetic field arising from the system *G* cannot penetrate the superconducting cylinder. When the current distribution in the system *G* changes, the shielding current distribution on the outer surface of the superconducting cylinder will change with it. So, the electrons in the system *G* and the electrons in the superconducting cylinder should be taken as one system, denoted as $\Sigma$. We assume that the wave function of the system $\Sigma$ is $|\varphi_n\rangle$, which includes the electrons in the system $G$ and the electrons in the superconducting cylinder. The Hamiltonian $\hat{H}$ of the system $\Sigma$ can be divided into two terms: $\hat{H} = \hat{H}_1 + \hat{H}_2$ where

$$\hat{H}_1 = \int dx^3 \hat{\psi}_1^+(\mathbf{x}) \frac{1}{2m} [-i\hbar\nabla + e\mathbf{A}(\mathbf{x})]^2 \hat{\psi}_1(\mathbf{x}) + \hat{V}_1 \qquad (5)$$

$$\hat{H}_2 = \int dx^3 \hat{\psi}_2^+(\mathbf{x}) \frac{1}{2m} [-i\hbar\nabla + e\mathbf{A}(\mathbf{x})]^2 \hat{\psi}_2(\mathbf{x}) + \hat{V}_2 \qquad (6)$$

Here, $\hat{H}_1$ represents the Hamiltonian of the electrons in the system $G$, $\hat{\psi}_1^+(\mathbf{x})$ and $\hat{\psi}_1(\mathbf{x})$ represent the creation and annihilation operators of these electrons respectively, and $\hat{V}_1$ denotes all the other potential terms in $G$ which has no relation to the magnetic field. $\hat{H}_2$ represents the Hamiltonian of the electrons in the superconducting cylinder, $\hat{\psi}_2^+(\mathbf{x})$ and $\hat{\psi}_2(\mathbf{x})$ represent the creation and annihilation operators of these electrons respectively, and $\hat{V}_2$ represents all the other potential terms in the cylinder, which has no relation to the magnetic field either, including the effective interaction between the superconducting electrons due to phonon exchange.

The vector potential $\mathbf{A}$ in exp(5) and exp(6) includes two parts $\mathbf{A} = \mathbf{A}_0 + \mathbf{A}_1$. The first vector potential $\mathbf{A}_0$ is defined by the static magnetic field $\mathbf{B}_0$, i.e. $\nabla \times \mathbf{A}_0 = \mathbf{B}_0$, where $\mathbf{B}_0$ is created by the solenoid, and only exists in the region enclosed by the solenoid. The second vector potential $\mathbf{A}_1$ is defined by the magnetic field $\mathbf{B}_1$, i.e. $\nabla \times \mathbf{A}_1 = \mathbf{B}_1$, where $\mathbf{B}_1$ is produced by both the currents in the system $G$ and on the outer surface of the superconducting cylinder, and only exists outside the cylinder.

When the superconductor cylinder comes into the superconducting state, the magnetic flux enclosed by it should be quantized in units of $h/2e$. So, for simplicity, we assume that the magnetic flux $\Phi$ created by the solenoid equals to $nh/2e$. In this case, when the superconductor cylinder comes into the superconducting state, no current is formed on its inner surface, and only the currents on its outer surface need to be taken into account. Therefore, the

Hamiltonian of the system $\Sigma$ can be rewritten as:

$$\hat{H} = \hat{H}_1 + \hat{H}_2$$
$$= \int dx^3 \hat{\psi}_1^+(\mathbf{x}) \frac{1}{2m} \{-i\hbar\nabla + e[\mathbf{A}_0(\mathbf{x}) + \mathbf{A}_1(\mathbf{x})]\}^2 \hat{\psi}_1(\mathbf{x}) + \hat{V}_1 \quad (7)$$
$$+ \int dx^3 \hat{\psi}_2^+(\mathbf{x}) \frac{1}{2m} \{-i\hbar\nabla + e[\mathbf{A}_0(\mathbf{x}) + \mathbf{A}_1(\mathbf{x})]\}^2 \hat{\psi}_2(\mathbf{x}) + \hat{V}_2$$

According to the variational method, the wave function $|\varphi_n\rangle$ satisfies the following variational equation $\delta\bar{H} = \delta\langle\varphi_n|\hat{H}|\varphi_n\rangle = 0$ under the normalized condition $\langle\varphi_n|\varphi_n\rangle = 1$. Besides this boundary condition, the Meissner effect provides another boundary condition. For the magnetic field $\mathbf{B}_1(\mathbf{x})$ only exists in the region outside the cylinder and cannot superposition with the magnetic field $\mathbf{B}_0(\mathbf{x})$ enclosed by the solenoid, the superimposed energy $W'$ between $\mathbf{B}_1(\mathbf{x})$ and $\mathbf{B}_0(\mathbf{x})$ must be zero:

$$W' = \int \frac{1}{\mu_0} \mathbf{B}_0(\mathbf{x}) \cdot \mathbf{B}_1(\mathbf{x}) dx^3 = \int \mathbf{A}_0(\mathbf{x}) \cdot \mathbf{J}(\mathbf{x}) dx^3 = 0 \quad (8)$$

where, $\mathbf{J}(\mathbf{x})$ is the current density, including the current density $\mathbf{J}_1(\mathbf{x})$ in the system $G$ and the shielding current density $\mathbf{J}_2(\mathbf{x})$ on the outer surface of the superconducting cylinder. $\mathbf{J}(\mathbf{x})$ can be rewritten as:

$$\mathbf{J}(\mathbf{x}) = \mathbf{J}_1(\mathbf{x}) + \mathbf{J}_2(\mathbf{x})$$
$$= \langle\varphi_n| \left\{ \frac{ie\hbar}{2m} \left[ \hat{\psi}_1^+(\mathbf{x})\nabla\hat{\psi}_1(\mathbf{x}) - \left(\nabla\hat{\psi}_1^+(\mathbf{x})\right)\hat{\psi}_1(\mathbf{x}) \right] - \frac{e^2}{m}[\mathbf{A}_0(\mathbf{x}) + \mathbf{A}_1(\mathbf{x})]\hat{\psi}_1^+(\mathbf{x})\hat{\psi}_1(\mathbf{x}) \right.$$
$$\left. + \frac{ie\hbar}{2m} \left[ \hat{\psi}_2^+(\mathbf{x})\nabla\hat{\psi}_2(\mathbf{x}) - \left(\nabla\hat{\psi}_2^+(\mathbf{x})\right)\hat{\psi}_2(\mathbf{x}) \right] - \frac{e^2}{m}[\mathbf{A}_0(\mathbf{x}) + \mathbf{A}_1(\mathbf{x})]\hat{\psi}_2^+(\mathbf{x})\hat{\psi}_2(\mathbf{x}) \right\} |\varphi_n\rangle$$

(9)

Therefore,

$$W' = \int \mathbf{A}_0(\mathbf{x}) \cdot \mathbf{J}(\mathbf{x}) dx^3$$

$$= \int dx^3 \mathbf{A}_0(\mathbf{x}) \cdot \langle \varphi_n | \left\{ \frac{ie\hbar}{2m} \left[ \hat{\psi}_1^+(\mathbf{x}) \nabla \hat{\psi}_1(\mathbf{x}) - \left( \nabla \hat{\psi}_1^+(\mathbf{x}) \right) \hat{\psi}_1(\mathbf{x}) \right] - \frac{e^2}{m} [\mathbf{A}_0(\mathbf{x}) + \mathbf{A}_1(\mathbf{x})] \hat{\psi}_1^+(\mathbf{x}) \hat{\psi}_1(\mathbf{x}) \right.$$
$$\left. + \frac{ie\hbar}{2m} \left[ \hat{\psi}_2^+(\mathbf{x}) \nabla \hat{\psi}_2(\mathbf{x}) - \left( \nabla \hat{\psi}_2^+(\mathbf{x}) \right) \hat{\psi}_2(\mathbf{x}) \right] - \frac{e^2}{m} [\mathbf{A}_0(\mathbf{x}) + \mathbf{A}_1(\mathbf{x})] \hat{\psi}_2^+(\mathbf{x}) \hat{\psi}_2(\mathbf{x}) \right\} | \varphi_n \rangle$$

(10)

Using the integration by parts, and noticing $\hat{\psi}_1^+(\mathbf{x}), \hat{\psi}_1(\mathbf{x}), \hat{\psi}_2^+(\mathbf{x}), \hat{\psi}_2(\mathbf{x}) \to 0$ at $x \to \infty$ we have:

$$W' = \langle \varphi_n | \int dx^3 \mathbf{A}_0(\mathbf{x}) \cdot \left\{ \frac{ie\hbar}{m} \hat{\psi}_1^+(\mathbf{x}) \nabla \hat{\psi}_1(\mathbf{x}) - \frac{e^2}{m} [\mathbf{A}_0(\mathbf{x}) + \mathbf{A}_1(\mathbf{x})] \hat{\psi}_1^+(\mathbf{x}) \hat{\psi}_1(\mathbf{x}) \right.$$
$$\left. + \frac{ie\hbar}{m} \hat{\psi}_2^+(\mathbf{x}) \nabla \hat{\psi}_2(\mathbf{x}) - \frac{e^2}{m} [\mathbf{A}_0(\mathbf{x}) + \mathbf{A}_1(\mathbf{x})] \hat{\psi}_2^+(\mathbf{x}) \hat{\psi}_2(\mathbf{x}) \right\} | \varphi_n \rangle = 0$$

(11)

This is a bounded condition added by the Meissner effect, called as the second boundary condition .

Now, we deal with the variational equation $\delta \overline{H} = 0$:

$$\delta \overline{H} = \delta \langle \varphi_n | \hat{H} | \varphi_n \rangle = \delta \langle \varphi_n | \int dx^3 \hat{\psi}_1^+(\mathbf{x}) \frac{1}{2m} [-i\hbar \nabla + e\mathbf{A}_0(\mathbf{x}) + e\mathbf{A}_1(\mathbf{x})]^2 \hat{\psi}_1(\mathbf{x}) + \hat{V}_1$$
$$+ \int dx^3 \hat{\psi}_2^+(\mathbf{x}) \frac{1}{2m} [-i\hbar \nabla + e\mathbf{A}_0(\mathbf{x}) + e\mathbf{A}_1(\mathbf{x})]^2 \hat{\psi}_2(\mathbf{x}) + \hat{V}_2 | \varphi_n \rangle$$

(12)

Expanding the above equation, we obtain

$$\delta \overline{H} = \delta \langle \varphi_n | \int dx^3 \left\{ \hat{\psi}_1^+(\mathbf{x}) \frac{1}{2m} [-i\hbar \nabla + e\mathbf{A}_1(\mathbf{x})]^2 \hat{\psi}_1(\mathbf{x}) + \hat{V}_1 \right\}$$
$$+ \int dx^3 \left\{ \hat{\psi}_2^+(\mathbf{x}) \frac{1}{2m} [-i\hbar \nabla + e\mathbf{A}_1(\mathbf{x})]^2 \hat{\psi}_2(\mathbf{x}) + \hat{V}_2 \right\} | \varphi_n \rangle$$
$$- \delta \langle \varphi_n | \int dx^3 \mathbf{A}_0(\mathbf{x}) \cdot \left\{ \frac{ie\hbar}{m} \hat{\psi}_1^+(\mathbf{x}) \nabla \hat{\psi}_1(\mathbf{x}) - \frac{e^2}{m} [\mathbf{A}_0(\mathbf{x}) + \mathbf{A}_1(\mathbf{x})] \hat{\psi}_1^+(\mathbf{x}) \hat{\psi}_1(\mathbf{x}) \right.$$
$$\left. + \frac{ie\hbar}{m} \hat{\psi}_2^+(\mathbf{x}) \nabla \hat{\psi}_2(\mathbf{x}) - \frac{e^2}{m} [\mathbf{A}_0(\mathbf{x}) + \mathbf{A}_1(\mathbf{x})] \hat{\psi}_2^+(\mathbf{x}) \hat{\psi}_2(\mathbf{x}) \right\} | \varphi_n \rangle$$
$$- \delta \langle \varphi_n | \int dx^3 \frac{e^2 \mathbf{A}_0^2(\mathbf{x})}{2m} [\hat{\psi}_1^+(\mathbf{x}) \hat{\psi}_1(\mathbf{x}) + \hat{\psi}_2^+(\mathbf{x}) \hat{\psi}_2(\mathbf{x})] | \varphi_n \rangle = 0$$

(13)

$\delta \overline{H}$ consists of three terms: The first term does not include the vector potential $\mathbf{A}_0(\mathbf{x})$; The

second term represents the superimposed energy between $\mathbf{B}_1(\mathbf{x})$ and $\mathbf{B}_0(\mathbf{x})$. According to the second boundary condition, this term must be zero, which is an inevitable result of the Meissner effect; The third term is the square term of $\mathbf{A}_0$ and can be neglected, for the vector potential $\mathbf{A}_0$ is a small quantity. Thus, we can obtain

$$\delta \bar{H} = \delta \langle \varphi_n | \hat{H} | \varphi_n \rangle = \delta \langle \varphi_n | \int dx^3 \left\{ \hat{\psi}_1^+(\mathbf{x}) \frac{1}{2m} [-i\hbar \nabla + e\mathbf{A}_1(\mathbf{x})]^2 \hat{\psi}_1(\mathbf{x}) + \hat{V}_1 \right\}$$
$$+ \int dx^3 \left\{ \hat{\psi}_2^+(\mathbf{x}) \frac{1}{2m} [-i\hbar \nabla + e\mathbf{A}_1(\mathbf{x})]^2 \hat{\psi}_2(\mathbf{x}) + \hat{V}_2 \right\} | \varphi_n \rangle = 0$$
(14)

In this equation, the vector potential $\mathbf{A}_0$ which describes the static magnetic field created by the solenoid has disappeared. It means that the magnetic field enclosed by the superconducting cylinder has no effect on the wave function of the system *G* which lies outside the cylinder. In fact, this variational equation is the same one which describes the case where the magnetic flux in the solenoid is zero. So, if the magnetic field is completely encircled by superconductors, the A-B effect will vanish outside the superconductors.

## 4   A new experimental scheme

This theorem demonstrates that the A-B effect does result from the superimposed energy between the magnetic fields. This conclusion should be proved by experiments. We know that the physical essence of the SQUID (Superconducting Quantum Interference Device[17] [18]) is similar to that of the A-B effect. The critical current $I_C$ passing through the SQUID depends on the flux $\Phi$ enclosed by the superconducting loop,

$$I_C = I_0 \left| \cos\left(\frac{\pi \Phi}{\Phi_0}\right) \right| \tag{15}$$

Where $\Phi_0 = h/2e$, for the current carriers in the SQUID are Cooper pairs. Our experimental scheme is showed in Fig. 2. A SQUID with two Josephson Junctions is placed in the plane of $z = 0$. A magnetic field $\mathbf{B}_0(\mathbf{x})$ confined by a long and straight solenoid is perpendicular to the plane and passes through the region enclosed by the superconducting loop. The vector

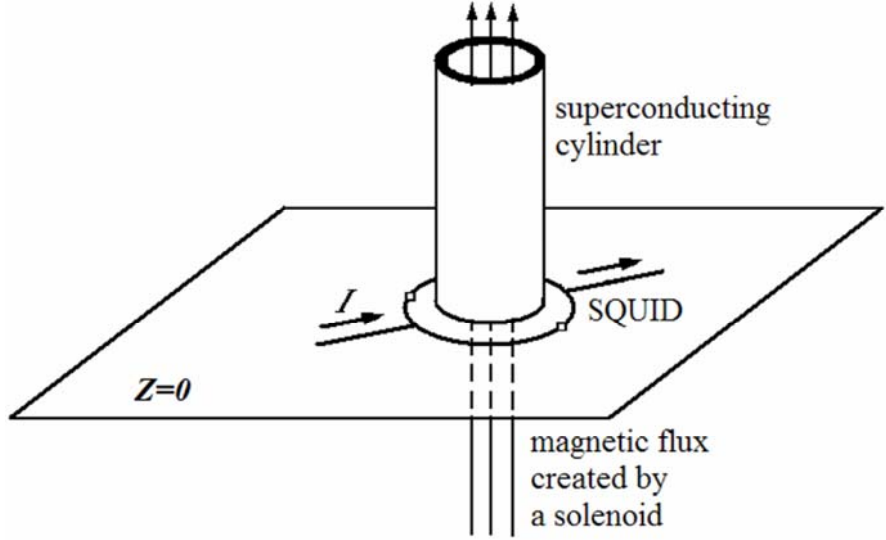

**Fig2,** A new experimental scheme which can be used to test the dynmatical cause of the A-B effect.

potential due to this confined flux is denoted by $\mathbf{A}_0$. The solenoid is divided into two parts by the plane of $z = 0$: The part above the plane is coated by a superconductor cylinder and the part below the plane is naked. Therefore, the magnetic field produced by the currents in the SQUID can only superposition with the magnetic field $\mathbf{B}_0(\mathbf{x})$ below the plane, with the superimposed energy being equal to $\int \frac{1}{2} \mathbf{A}_0(\mathbf{x}) \cdot \mathbf{J}'(\mathbf{x}) dx^3$ instead of $\int \mathbf{A}_0(\mathbf{x}) \cdot \mathbf{J}'(\mathbf{x}) dx^3$, where $\mathbf{J}'(\mathbf{x})$ denotes the current density in the SQUID.

So, if the A-B effect results from the superimposed energy between the magnetic fields, the critical $I_C$ passing through the SQUID should be determined by the following equation

$$I_C = I_0 \left| \cos\left[ \frac{\pi(\Phi/2)}{\Phi_0} \right] \right| \tag{16}$$

When $\Phi = 2n\Phi_0$, $I_C = I_0$; When $\Phi = (2n+1)\Phi_0$, $I_C = 0$.

If the A-B effect results from the vector potential $\mathbf{A}_0$, but not the superimposed energy between the fields, the critical current $I_C$ should still be determined by the exp (15). No matter for $\Phi = 2n\Phi_0$ or for $\Phi = (2n+1)\Phi_0$ both the critical current $I_C$ should be $I_0$.

Therefore, this experimental scheme can provide us a direct method to judge that the dynamical mechanism of the A-B effect is due to the superimposed energy between the magnetic fields? or due to the existence of the vector potential $\mathbf{A}_0$ ?

# 5  Conclusion

In classical physics, the motion of a particle is determined by the forces acting on it. The force is a local conception, which is determined by some local quantities, such as $\mathbf{B}(\mathbf{x})$ or $\mathbf{E}(\mathbf{x})$. If a particle doesn't pass through the region where the magnetic field exists, the magnetic field cannot affect the motion of the particle, for it cannot exert any forces on the particle. But in quantum mechanics, the motion of a particle is determined by the Hamiltonian, which represents the energy of the system. The energy of the system is a global concept, which is determined by the distribution of the electromagnetic field in the whole space and is related to some global quantities, such as the vector potential $\mathbf{A}(\mathbf{x})$ or the scale potential $\phi(\mathbf{x})$. In quantum physics, even though a particle only passes through the region where the static magnetic field does not exist, the particle can still be affected by this static magnetic field, for the magnetic field produced by the moving particle can possibly superposition with this static magnetic field and cause the change of the system's energy. Just for this reason, the A-B effect was observed in the experiments.

# 6  Acknowledgement

The author thanks Professors C. N. Yang, R. S. Han, J. M. Li for their helpful discussions. This work is supported by Science and Technology Foundation of Beijing Jiaotong University (Grant No. 2005sm058)